\documentclass[aps,prx,twocolumn,notitlepage,showpacs,amsmath,amstex,amssymb,citeautoscript,longbibliography]{revtex4-1}
\pdfoutput=1
\usepackage{natbib}
\usepackage[english]{babel}
\usepackage{letltxmacro}
\usepackage{latexsym}
\LetLtxMacro{\ORIGselectlanguage}{\selectlanguage}
\makeatletter
\DeclareRobustCommand{\selectlanguage}[1]{%
  \@ifundefined{alias@\string#1}
    {\ORIGselectlanguage{#1}}
    {\begingroup\edef\x{\endgroup
       \noexpand\ORIGselectlanguage{\@nameuse{alias@#1}}}\x}%
}
\newcommand{\definelanguagealias}[2]{%
  \@namedef{alias@#1}{#2}%
}
\makeatother
\definelanguagealias{en}{english}
\definelanguagealias{English}{english}
\usepackage{graphicx}
\usepackage{amsmath}
\usepackage{amsfonts}
\usepackage{amssymb,bbding}
\usepackage{bm}
\usepackage[title]{appendix}
\usepackage{color}
\usepackage[percent]{overpic}
\usepackage{soul} 
\usepackage{amssymb}
\usepackage{wasysym}
\usepackage{dsfont}
\usepackage{float}
\usepackage{braket}
\usepackage{cancel}
\usepackage{comment}
\usepackage{hyperref}
\hypersetup{
    bookmarks=false,         
    unicode=false,          
    pdftoolbar=false,        
    pdfmenubar=true,        
    pdffitwindow=false,     
    pdfstartview={FitH},    
    pdftitle={Stabilizing two-dimensional quantum scars by deformation and synchronization},    
    pdfauthor={A. A. Michailidis, C. J. Turner, Z. Papic, D. A. Abanin, and M. Serbyn},     
    pdfsubject={},   
    pdfcreator={},   
    pdfproducer={}, 
    pdfkeywords={quantum scars} {non-ergodic dynamics}{mixed phase space}{quantum many-body chaos}{thermalization}, 
    pdfnewwindow=true,      
    colorlinks=true,       
    linkcolor=black,          
    citecolor=blue,        
    filecolor=magenta,      
    urlcolor=blue           
}
\newcommand{\norm}[1]{\left\lVert#1\right\rVert}

\newcommand{\be}{\begin{equation}}
\newcommand{\ee}{\end{equation}}
\newcommand{\bea}{\begin{eqnarray}}
\newcommand{\eea}{\end{eqnarray}}

\renewcommand{\vec}[1]{{\bf #1}}

\begin{document}
\title{Stabilizing two-dimensional quantum scars by deformation and synchronization} 
\author{A. A. Michailidis$^1$, C. J. Turner$^2$, Z. Papi\'c$^2$, D. A. Abanin$^3$, and M. Serbyn$^1$}
\affiliation{$^1$IST Austria, Am Campus 1, 3400 Klosterneuburg, Austria}
\affiliation{$^2$School of Physics and Astronomy, University of Leeds, Leeds LS2 9JT, United Kingdom}
\affiliation{$^3$Department of Theoretical Physics, University of Geneva, 24 quai Ernest-Ansermet, 1211 Geneva, Switzerland}

\date{\today}
\begin{abstract}
Relaxation to a  thermal state is the inevitable fate of non-equilibrium interacting quantum systems without special conservation laws. While thermalization in one-dimensional (1D) systems can often be suppressed by integrability mechanisms, in two spatial dimensions thermalization is expected to be far more effective due to the increased phase space. In this work we propose a general framework for escaping or delaying the emergence of the thermal state in two-dimensional (2D) arrays of Rydberg atoms via the mechanism of quantum scars, i.e.~initial states that fail to thermalize. The suppression of thermalization is achieved in two complementary ways: by adding local perturbations or by adjusting the driving Rabi frequency according to the local connectivity of the lattice.  We demonstrate that these mechanisms allow to realize robust quantum scars in various two-dimensional lattices, including decorated lattices with non-constant connectivity.  In particular, we show that a small decrease of the Rabi frequency at the corners of the lattice is crucial for mitigating the strong boundary effects in two-dimensional systems. Our results identify synchronization as an important tool for future experiments on two-dimensional quantum scars. 
\end{abstract}
\maketitle
 
{\it Introduction.}---Recent experimental breakthroughs allow to probe non-equilibrium quantum dynamics of various isolated quantum systems~\citep{lewenstein2012ultracold,blatt2012quantum,browaeys2020many}. Yet, for generic interacting systems that do not have any special conservation laws, such dynamics leads to a thermal state. This process of thermalization is explained by the typicality of highly excited eigenstates in interacting quantum systems. Formally, the Eigenstate Thermalization Hypothesis (ETH)~\citep{DeutschETH,SrednickiETH} conjectures that all eigenstates of a Hamiltonian in a sufficiently narrow energy shell, display the same expectation values of physical observables as the microcanonical ensemble. ETH has been numerically and experimentally verified in a variety of different quantum systems~\citep{Deutsch_2018,Polkovnikov-rev}. 

In order to observe long-time coherent dynamics in quantum systems one must avoid thermalization or at least delay its onset. Integrable systems which satisfy the Yang-Baxter equation~\citep{bethe1931theorie,faddeev1996algebraic}, and the disordered systems which undergo a many-body localization (MBL) transition~\citep{Basko06,RevModPhys.91.021001}, provide explicit examples of ETH violation. However,  integrability is known to exist only for 1D systems; the existence of MBL in higher dimensions is also debated~\cite{Roeck-MBL2d,Choi}. Intuitively, thermalization is more ubiquitous in higher dimensions due to larger phase space available for relaxation processes. This motivates the exploration of alternative ETH-violating mechanisms. 

Recent experiments on Rydberg atom arrays~\cite{Bernien2017} suggested the possibility of  \emph{weak} ETH breaking via a different mechanism now known as ``quantum many-body scars"~\citep{Turner2017,wenwei18}. Quantum many-body scarring manifests itself as the presence of a small set of atypical ETH-breaking eigenstates. Experimentally, scars lead to strong dependence of relaxation on initial conditions: initial configurations that have a large overlap with atypical eigenstates feature slow growth of entanglement and long-time coherent dynamics, whereas other initial states relax much faster. Theoretically, scars have been explained via the existence of an (un)stable trajectory within the variational semiclassical approach~\cite{wenwei18,michailidis2019slow} or, alternatively, via a hidden $\mathrm{su(2)}$ algebra representation in the subspace of atypical eigenstates~\cite{PhysRevLett.122.220603, Bull2020}. In addition, some exact scarred eigenstates of the Rydberg atom chain have been constructed~\cite{lin2018exact}, and their stability under perturbations was investigated~\cite{Khemani2018,lin2019}. Finally, scars were also reported in a variety of other models~\cite{Vafek, Moudgalya2018, Schecter2019, IadecolaZnidaric, Bull2019, NeupertScars, mukherjee2019collapse,
Haldar2019, sugiura2019many,ho19p, Iadecola2019_3, Hudomal2019, Zhao2020}, while scarring may be related to non-ergodic behavior observed in models with confinement~\cite{Calabrese16, Konik1, Konik2}, dynamical symmetries~\cite{Buca2019,Tindall2019}, fractons~\cite{Pretko19, Khemani2019, Sala2019,  Khemani2019_2}, and ``Krylov restricted thermalization"~\cite{MoudgalyaKrylov}.

\begin{figure*}[t]
\begin{center}
\includegraphics[width=1.99\columnwidth]{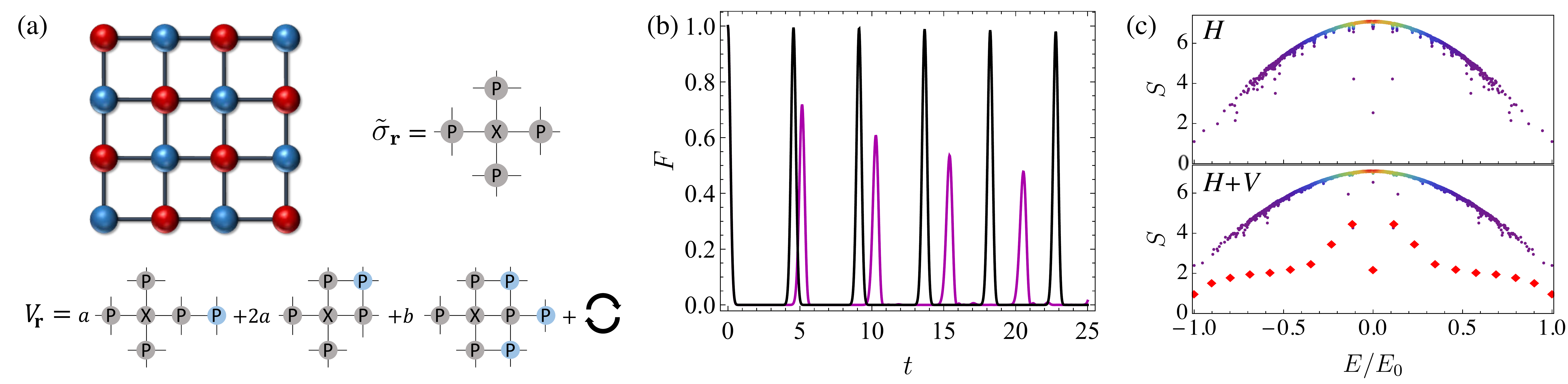}
\caption{ \label{Fig:definitions}  (a) Square lattice, Hamiltonian density operator, Eq.~(\ref{Eq:Ham}) and the deformation, Eq.~(\ref{Eq:pert}) needed to stabilize the scars. (b) Fidelity of quantum many-body revivals for the unperturbed Hamiltonian (magenta) and the model with the optimal perturbation (black) for $6\times 6$ lattice with PBC. (c) Entanglement entropy for all eigenstates as a function of their energy. The color indicates the density of dots, which is strongly peaked around zero energy. The deformed model (bottom plot) has a much more pronounced band of low-entangled eigenstates (red diamonds) compared to the underformed  model (top plot). Data is obtained by exact diagonalization in the zero-momentum, inversion-symmetric sector for $6\times 6$ lattice with $\mathcal{ D} = 9702 $ states. }
\end{center}
\end{figure*}

In this work we present a detailed study of scars on 2D lattices of Rydberg atoms in the regime of the nearest-neighbor blockade that has been realized in many recent experiments~\cite{Labuhn2016,Bernien2017,PhysRevX.4.021034,Browaeys2020}. We concentrate on experimental knobs that could be used to enhance many-body scars in 2D quantum systems, which are significantly more susceptible to thermalization as well as finite-size effects due to their larger boundary-to-bulk ratio. First, we show that weak perturbations of the Rydberg atom Hamiltonian on square lattices can significantly stabilize scars by improving an approximate $\mathrm{su(2)}$ algebra representation in the subspace of scarred eigenstates. This leads to stronger fidelity revivals and enhanced coherence in the dynamics. Further, we consider scars on more complicated lattices and in the presence of open boundaries. For lattices featuring non-uniform connectivity, coherent many-body oscillations can be stabilized by adjusting the driving Rabi frequency according to local connectivity. We refer to this stabilization mechanism as ``enforced synchronization", and we demonstrate that this can be used to suppress the dephasing due to the boundary by matching the oscillation frequency at the boundary and in the bulk.  

\emph{Model.}---We begin by considering Rydberg atoms arranged in a square lattice in the regime of the nearest-neighbor blockade. The Hamiltonian generates Rabi oscillations of a given atom under the constraint that all four neighboring atoms are in the ground state, 
\begin{equation}\label{Eq:Ham}
H = \sum_{\vec{r}}\sigma^x_{\vec{r}}\prod_{\braket{\vec{r'},\vec{r}}} P_{\vec{r'}} = \sum_{\vec{r}}{\tilde \sigma}^x_{\vec{r}}, 
\end{equation}
where the indices, $\vec{r} = (i,j)$, denote  the lattice site, $i, j=1,\ldots L$, and the product goes over all nearest neighbors of site $\vec{r}$. The operator $\sigma^x_{\vec{r}} =  \ket{\uparrow}\bra{\downarrow}+ \ket{\downarrow}\bra{\uparrow}$ describes Rabi oscillations between excited ($\uparrow$) and ground states ($\downarrow$) of a given atom. The product of projectors onto the ground state, ${P = \ket{\downarrow}\bra{\downarrow}}$, ensures the absence of excitations on nearest neighbor sites.  In Fig.~\ref{Fig:definitions}(a) we show the lattice and the corresponding Hamiltonian density operator, ${\tilde \sigma}^x_{\vec{r}}$. We focus on the sector of the Hilbert space with no adjacent excitations, which is the largest sector of the system. The dimension of this sector scales as  $\text{Dim}(\mathcal{H}) \propto c_1^{L^2}$ where  $c_1 \approx 1.503...$ is the hard square entropy constant~\citep{Baxter1999}.

\emph{Stabilization of scars via deformation.}---Figure~\ref{Fig:definitions}(a) shows a partition of the square lattice $\mathcal{M}$ into two sublattices, $\mathcal{M} = A \cup B$. Two states with the maximum number of excitations (compatible with the constraint of no adjacent excitations), $\ket{M_{A (B)}}$, correspond to all the atoms in sublattice $A$ ($B$) being in the excited state. In Ref.~\citep{michailidis2019slow}, it was shown that the fidelity, $F(t)=|\langle M_A|e^{-iHt}|M_A \rangle|^2$, which quantifies a probability of returning to the many-body state $\ket{M_A}$ at time $t$ features persistent revivals with period $T$. These revivals were attributed to the existence of a periodic trajectory in the variational manifold of tree tensor states.

Figure~\ref{Fig:definitions}(b) shows the revivals for a $6 \times 6$ square lattice with periodic boundary conditions (PBC).  The persisting  oscillations of fidelity have a period of  $T \approx 5$, where at half-period the system is approximately close to the second maximally excited state, $\ket{M_B}$. This dynamics is similar to the 1D case where the system oscillates between the two Ne\'el states~\cite{Bernien2017}. The revivals are decaying, and it is interesting to find small deformations that would enhance them.

In order to improve the revival quality, we propose the following deformation of the Hamiltonian, see Fig.~\ref{Fig:definitions}(a),
\begin{equation}\label{Eq:pert}
V = \sum_{\vec{r}}V_{\vec{r}},\quad V_{\vec{r}}  = {\tilde\sigma}^x_{\vec{r}}(a \mathcal{P}_{\vec{r}}^{l}+ 2 a \mathcal{P}_{\vec{r}}^{d}+ b\mathcal{P}_{\vec{r}}^{3}),
\end{equation}
where $a$ and $b$ are parameters to be optimized and the projectors are defined as
\begin{subequations}\label{Eq:perturbations}
\begin{eqnarray}
\mathcal{P}^{l}_{i,j} &=& P_{i,j+2}+ \ldots,\\
\mathcal{P}^{d}_{i,j} &=& P_{i+1,j+1}+ \ldots,\\
\mathcal{P}^{3}_{i,j} &=& P_{i-1,j+1}P_{i,j+2}P_{i+1,j+1}+ \ldots.
\end{eqnarray}
\end{subequations}
Ellipses in Eqs.~(\ref{Eq:perturbations}) denote the three remaining terms obtained by $90^\circ$ rotations around the lattice site at position $\vec{r}=(i,j)$ that make the perturbation invariant under the full space group symmetry. Our heuristics on the choice of perturbations are based on the ``forward scattering approximation" (FSA)~\citep{Turner2017,Turner2018,PhysRevLett.122.220603}. Intuitively, the three terms in the deformation~(\ref{Eq:pert}) correspond to configurations encountered in the process of flipping the four excited Rydberg atoms that are nearest neighbors on the $A$ sublattice into their ground state~\cite{SOM}. 

Optimization of coefficients $a$ and $b$ for the $6 \times 6$ size lattice with PBC results in $a \approx 0.040$, $b \approx 0.056$. The optimization of  $a,b$ is performed by maximizing the fidelity at the first revival, $F(T)$, using the Nelder-Mead method, see~\cite{SOM}. The resulting fidelity time series are shown in  Fig.~\ref{Fig:definitions}(b) where one observes a significant improvement of the revival quality from $F \approx 0.72$ for the unperturbed model to $F \approx 0.997$ for the optimal perturbation.  

\emph{Structure of eigenstates.}---%
The effect of optimal deformation is strongly pronounced not only in the dynamics, but also in eigenstate properties, such as entanglement entropy.  Fig.~\ref{Fig:definitions}(c) compares the entanglement of each eigenstate for the clean and perturbed models. The entanglement is calculated as ${S= -\text{Tr}\{\rho_{\mathcal{L}}\log \rho_{\mathcal{L}}\}}$, where $\rho_{\mathcal{L}}= \text{Tr}_{\mathcal{R}}\ket{\psi}\bra{\psi}$  is the reduced density matrix for the bipartition of lattice into two cylindrical subsystems $\cal R, L$ of size $({L}/{2})\times L$, where $L$ is the linear dimension if the lattice. In both cases, the entropy for the majority of the eigenstates depends only on energy density, consistent with ETH.  The unperturbed system features no significant entanglement outliers, in contrast to 1D models where a similar plot clearly revealed the special scarred eigenstates~\cite{Turner2017,Turner2018}. At the same time, the special eigenstates still can be detected by their overlap with the $\ket{M_{A,B}}$ product states~\cite{SOM}.  By contrast, the optimally perturbed Hamiltonian displays a special band of eigenstates with much lower entropy than any other eigenstate at similar energy density, as seen in bottom panel of Fig.~\ref{Fig:definitions}(c). Likewise, the deformation enhances the overlap of special eigenstates with $\ket{M_{A,B}}$ product states. 

The existence of a deformation that improves the special band of eigenstates suggests that potentially one may deform the 2D Hamiltonian~(\ref{Eq:Ham}) to the point where the manifold of scarred eigenstates forms an exact $\mathrm{su(2)}$ representation. However, while Ref.~\cite{PhysRevLett.122.220603} provided strong numerical evidence for the existence of exact scars in 1D models by constructing a long-range quasi-local deformation, the rapidly growing Hilbert space of 2D systems precludes us from simulating longer range deformation terms. At the same time, the existence of such a perturbation in the 2D case is non-trivial and suggests that the existence of exact scars is not related to integrability~\cite{Khemani2018}.  Moreover, the leading order deformation  improves the coherence so strongly that longer range terms may be not needed on the experimentally relevant timescales. 

\begin{figure}[t]
\begin{center}
\includegraphics[width=0.99\columnwidth]{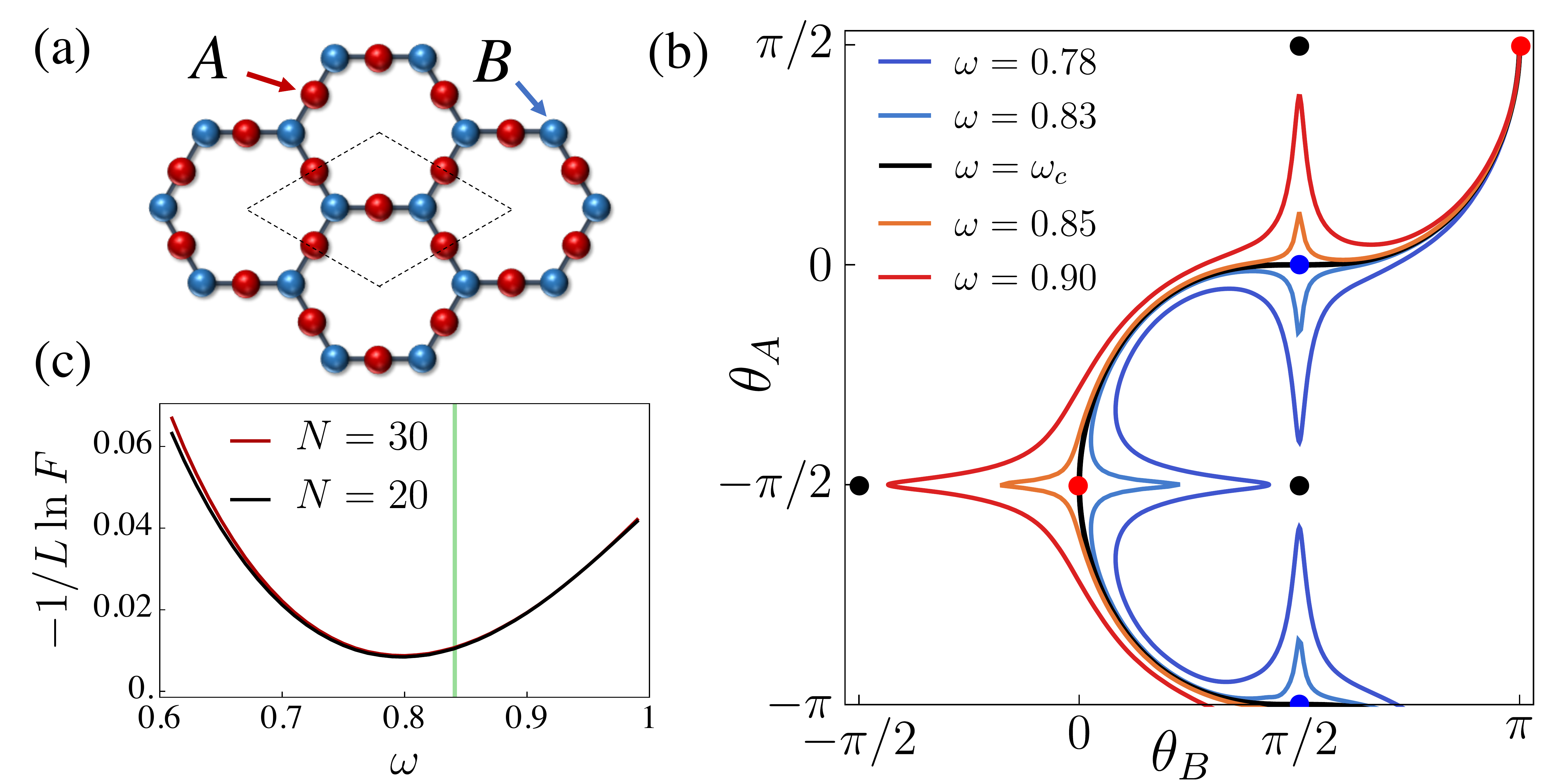}
\caption{ \label{Fig:decor}(a) The decorated honeycomb lattice where each site from $A$ ($B$) partition has 2 (3) neighbors. The unit cell contains five lattice cites.  (b) Plot of the TDVP trajectories for different $\omega$ and regularization $\epsilon  = 4 \cdot 10^{-4}$. Black points indicate the singular points, whereas red/blue points  correspond to  $\ket{M_{A/B}}$ states. (c) The fidelity of the first revival of the quantum Hamiltonian with respect to the frequency $\omega$  for two different lattice sizes $N$. The TDVP prediction for the optimal value of $\omega$ is illustrated by the green line. 
}
\end{center}
\end{figure}
\emph{Scars in decorated lattices.}---Above we considered the square lattice -- the simplest 2D bipartite lattice (see~\cite{SOM} for the case of honeycomb lattice). It is interesting to explore more exotic lattices, e.g., one possibility, which does not exist in 1D, are lattices where different Rydberg atoms have different number of nearest neighbors. 

The simplest bipartite lattice with different connectivity can be obtained from the honeycomb lattice by adding extra Rydberg atoms to the middle of each link, see Fig.~\ref{Fig:decor}(a). Such ``decorated'' honeycomb lattice is bipartite, where partition $A$ consists of atoms in the middle of the edges  and partition $B$ includes atoms located at the vertices of the honeycomb lattice. We assume there is no Rydberg blockade between sites on the same sublattice. Under such an assumption, we write the Hamiltonian as 
\begin{equation}\label{Eq:decor2}
H= \omega_{A} \sum_{\vec{r}\in A}{\tilde \sigma}^x_{\vec{r}}+ \omega_B \sum_{\vec{r} \in B}{\tilde \sigma}^x_{\vec{r}},
\end{equation}
where the Hamiltonian density operator is the same as in Eq.~(\ref{Eq:Ham}), and we introduced two different Rabi frequencies (we set $\omega_B=1$ for simplicity). We will tune $\omega_A$ below to correct for the connectivity mismatch between different sublattices. For the rest we denote $\omega_{A} \equiv \omega$, and use PBC. The maximally blocked states, $\ket{M_{A (B)}}$ are given by exciting every site from sublattice $A (B)$, while keeping the atoms in the other sublattice in their ground state. Now these states have inequivalent number of excited Rydberg atoms, with the ``maximally excited'' state in the system being $\ket{M_{A}}$.

To have a quantitative understanding of dynamics, we approximate the decorated lattice by a tree with the same pattern of local connectivities using the method discussed in Ref.~\citep{michailidis2019slow}. We project quantum dynamics on the tree onto a manifold of tensor tree states (TTS),  parametrized by two real angles $\ket{\psi(\theta_{A},\theta_{B}})$ using the time-dependent variational principle (TDVP)~\citep{kramer1981geometry, michailidis2019slow}~\cite{SOM}. The resulting equations of motion in the TTS manifold read,
\begin{subequations}\label{Eq:EOM}
\begin{eqnarray}
 \dot\theta_A & =&  -\omega  \cos^{c_{A}-1}{\theta_{B}}-\cos^{c_{B}}{\theta_{A}} \sin{\theta_{A}}   \tan{\theta_{B}} ,\\ 
  \dot\theta_B & =& -  \cos^{c_{B}-1}{\theta_{A}} - \omega \cos^{c_{A}}{\theta_{B}}\sin{\theta_{B}}   \tan{\theta_{A}},
\end{eqnarray}
\end{subequations}
where $c_{A} = 2$, $c_{B} = 3$ are the connectivities of sublattices $A,B$. For the case when $c_A=c_B$ and $\omega=1$, Refs.~\cite{michailidis2019slow,wenwei18} demonstrated the existence of a periodic trajectory that connects states $\ket{M_{A,B}}$ on the variational manifold. 

Surprisingly, when $c_A\neq c_B$ as in the present case, the trajectory emanating from the $\ket{M_A}$ state does not reach $\ket{M_B}$ state but instead falls down into the singular point. Thus, an unstable periodic orbit does not exist for generic values of $\omega$. In order for it to exist, it should pass through \emph{both} $\ket{M_{A}}$ and $\ket{M_{B}}$ states. Fig.~\ref{Fig:decor}(b) illustrates that this happens for a special value of the frequency, $\omega_{c}\approx 0.841$. Note that in this figure we regularized the equations of motion by replacing $\tan \theta_{A,B} \rightarrow \tan\theta_{A,B}/(1-\epsilon \tan^2\theta_{A,B})$, where the value of $\epsilon$ is small but finite. Such a regularization prevents trajectories from completely ``falling'' into singular points, yet we see that only at $\omega_c$ the trajectory passes through both $\ket{M_{A,B}}$ states, with the value of $\omega_c$ being independent of regularization. 

Finally, we investigate the behavior of quantum fidelity at the first revival as a function of $\omega$. Fig.~\ref{Fig:decor}(c) shows that the fidelity has best revivals at the value of $\omega \approx 0.8$, which is close to but does not  coincide  with the prediction from TDVP dynamics, $\omega_c$.  The difference between the two values and also the smooth dependence of fidelity revival quality on $\omega$ may be attributed to quantum fluctuations present in the model. 

The improvement of oscillations predicted by variational dynamics and confirmed in the simulation of exact quantum dynamics may be intuitively explained as enforced synchronization. Indeed, in the decorated honeycomb lattice the atoms on sublattice $A$ experience weaker blockade due to presence of a smaller number of nearest neighbors. Thus, the optimal fidelity revivals are achieved when Rabi frequency $\omega$ on this sublattice is decreased compared to sublattice $B$. We believe that such intuition will also hold for more decorated lattices with different local connectivities $c_A$ and $c_B$, see~\cite{SOM} for predictions for $\omega$ from FSA. On the one hand, this can open the door to the realization of scars on lattices with more exotic geometries; on the other hand, this intuition can be applied to remove the unwanted boundary effects, as we show next.

\begin{figure}[t]
\begin{center}
\includegraphics[width=0.99\columnwidth]{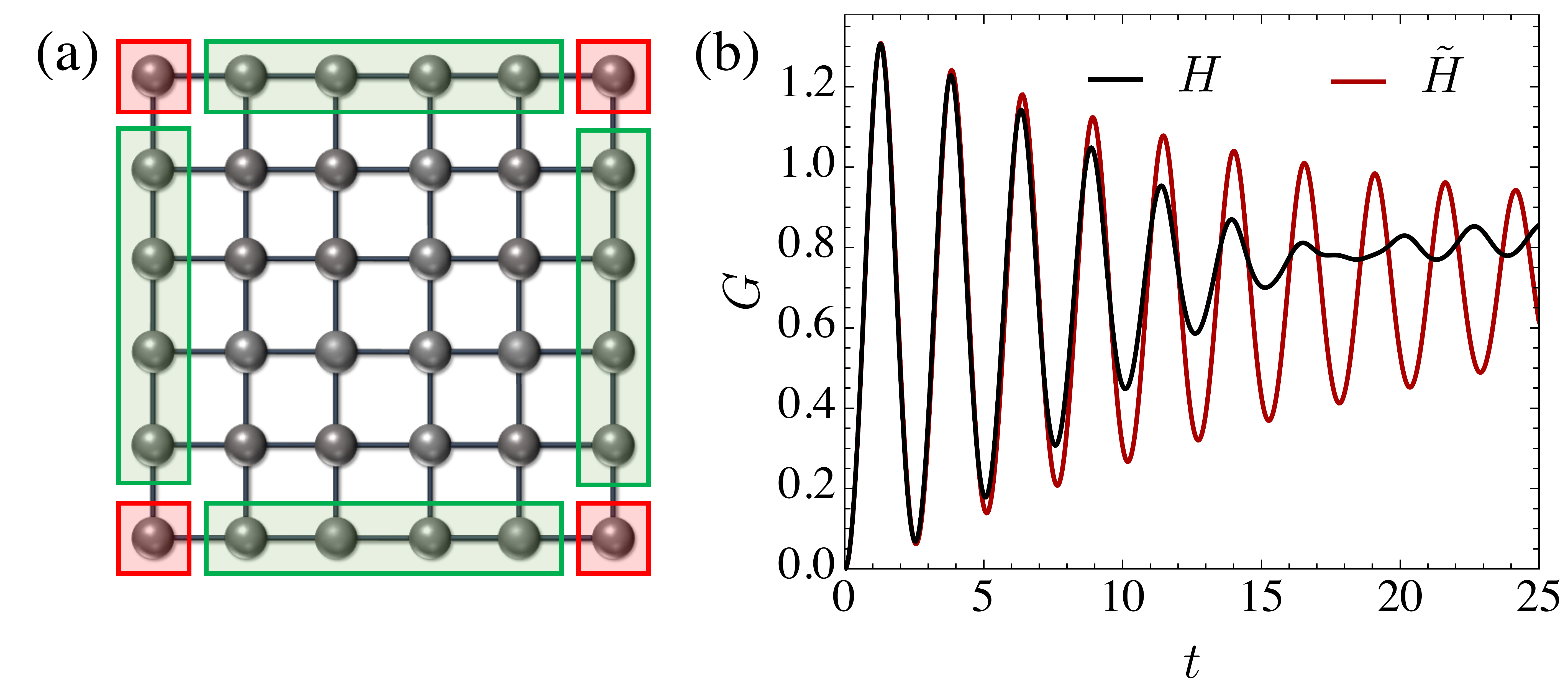}
\caption{ \label{Fig:open_fid}(a) Illustration of the splitting of the square lattice into three different regions distinguished by the number of nearest neighbors. (b) Domain wall dynamics for the maximally excited initial state of a $6 \times 6$ square lattice with \emph{open} boundary conditions. The black curve corresponds to the non-corrected Hamiltonian and the red curve corresponds to the system where the corner Rabi  frequency is reduced by $g_{c} = 0.105$.
}
\end{center}
\end{figure}

\emph{Boundary synchronization.}---
In experiments with Rydberg blockade, atoms are manipulated individually with optical tweezers~\citep{Labuhn2016,Bernien2017,PhysRevX.4.021034,Browaeys2020}, which enables realization of arbitrary lattice geometries.  At the same time, implementing PBC that were used above is challenging if not unfeasible. Thus it is imperative to understand and address boundary effects. For instance, the boundary for the square lattice as large as $6\times 6$ atoms still has more atoms compared to the ``bulk'' of the lattice -- see Fig.~\ref{Fig:open_fid}(a). Different number of local neighbors at the boundary and in the bulk of the system leads to faster dephasing that quickly degrades fidelity revivals as well as oscillations of local observables.  

Inspired by the results from decorated lattices, we propose a correction to the local Rabi frequency which depends on the local connectivity. The corrected Hamiltonian for the square lattice reads,
\begin{equation}
\tilde{H} = H - g_{C} \sum_{\vec{r}\in \mathcal{C}}{\tilde \sigma}^x_{\vec{r}} - g_{E} \sum_{\vec{r}\in \mathcal{E}} {\tilde \sigma}^x_{\vec{r}},
\end{equation}
where the $H$ is Hamiltonian from Eq.~(\ref{Eq:Ham}) and the subtracted terms include the sum over all atoms at corners ($\cal C$) which have only two nearest neighbors and those at the edges of lattice ($\cal E$), which have three neighbors, see Fig.~\ref{Fig:open_fid}(a).

To optimize the perturbations $(g_{C},g_{E})$, we maximize the fidelity on a $4 \times 4$ lattice where the full Hilbert space has dimension $\mathop{\text{Dim}}\mathcal{H} = 1234$. In this case we find an insignificant correction to the edge sites, $g_{E} \approx 10^{-3}$, while the corner terms acquire a much stronger correction, $g_{C} \approx 0.12$.  Guided by this result, we completely disregard the edge correction, by setting $g_{E} = 0$, and focus only on the correction to the four corners of the lattice, $g_{C}$. The optimization of fidelity for the $6\times 6$ lattice yields optimal value $g_{C}\approx 0.105$ which corresponds to an approximately $10\%$ decrease in the Rabi frequency for corners of the lattice. 

We explore the effects of the perturbation on the dynamics of the experimentally observable quantity --- mean domain wall density, 
$
G= ({1}/{L^2})\sum_{\vec{r}}P_{\vec{r}}\sum_{\braket{\vec{r'}\vec{r}}}P_{\vec{r'}}
$. Fig.~\ref{Fig:open_fid}(b) compares the dynamics of the domain wall density in the quench from $\ket{M_A}$ state for the original and boundary-synchronized  Hamiltonians with open boundary conditions. While at early times the effects of the boundaries are weak (the Lieb-Robinson bound~\cite{LiebRobinson} suggests that boundary effects ``propagate'' to the bulk with a constant velocity), after four revivals the dephasing from the boundaries begins to degrade the oscillations. For the uncorrected model the domain wall density is almost equilibrated at $t \gtrsim 15$. In contrast, the oscillations in the  synchronized Hamiltonian persist for much longer times. 

\emph{Discussion.}---%
We demonstrated the stabilization of quantum scars in 2D lattices by two complementary types of deformations of the Hamiltonian. First, we constructed a weak longer-range deformation that improves the quality of the fidelity revivals by further decoupling the scarred subspace away from the thermal bulk of the spectrum, similar to ``perfect" scars in 1D  Rydberg blockade~\citep{PhysRevLett.122.220603}. Second, inspired by TDPV description within the  TTS manifold~\cite{michailidis2019slow}, we proposed synchronization as a mechanism for improving scars on lattices of non-constant connectivity and in presence of boundaries. The local tuning of the Rabi frequency is feasible and can be used to experimentally mitigate the boundary effects. We expect that such a synchronization will open the door to the experimental application of scars in two dimensions akin to the $\pi$-pulse experiment in 1D~\citep{omran2019generation}.    

An immediate question raised by our results is the interplay between the synchronization mechanism explained via TDVP and the deformation of the Hamiltonian that is  explained in terms of $\mathrm{su(2)}$ representations. Understanding the relation between these two mechanisms beyond phenomenological arguments provided in~\cite{SOM} could provide a more complete picture and classification of possible scars. In addition, the existence of synchronization that improves scars bears a distant analogy to the collective oscillations in the BCS model~\cite{PhysRevLett.93.160401} and collective modes in Maxwell-Bloch equation~\cite{sit,*sit2}. Making this analogy more quantitative could prove fruitful for generalizations of scars.

More broadly, while we demonstrated the existence of scars for several bipartite lattices, the existence of oscillations in \emph{non-bipartite} lattices, such as triangular or kagome, remains an open question. For instance, triangular lattice features a natural partition into three sublattices and it would be interesting to explore the possibility for analogs of $Z_3$ scars in Rydberg chains~\cite{Turner2018,michailidis2019slow,Bull2019}. In addition, understanding the connection between existence of scars and ground state phase diagram~\cite{ho19} and extending these results to models with longer-range blockade remains an interesting question. 

\emph{Note added.}---When this work was at the final stages of preparation, Ref.~\cite{Lee20} proposed that XXZ spin-1/2 models may acquire non-thermal eigenstates on kagome lattice via a mechanism that utilizes geometric frustration. It remains to be understood if a similar mechanism could be useful for constrained models on non-bipartite lattices.

\emph{Acknowledgments.}---%
We acknowledge useful discussions with  W.~W.~Ho, H.~Pichler, S.~Choi, L.~Levitov, E.~Demler, and M.~Lukin.%

A.M. and M.S. were supported by the European Research Council (ERC) under the European Union's Horizon 2020 research and innovation programme (grant agreement No. 850899). C.J.T. and Z.P. acknowledge support by EPSRC grants EP/R020612/1 and EP/M50807X/1. Statement of compliance with EPSRC policy framework on research data: This publication is theoretical work that does not require supporting research data. DA acknowledges support by the Swiss National Science Foundation. This work benefited from participation at KITP Follow-Up program, supported by the National Science Foundation under Grant No. NSF PHY-1748958 and from the program ``Thermalization, Many body localization and Hydrodynamics'' at International Centre for Theoretical Sciences (Code: ICTS/hydrodynamics2019/11).

%

\clearpage
\pagebreak
\onecolumngrid
\begin{center}
\textbf{\large Supplementary material for ``Stabilizing two-dimensional quantum scars by deformation and synchronization'' }\\[5pt]
\begin{quote}
{\small 
In this supplement we present additional details on the FSA method and its relation to the su(2) representation for the square lattice. In addition, we illustrate the generalization of the deformation using the example of the honeycomb lattice. Finally we introduce the tensor tree ansatz used in the main text to approximate dynamics for the decorated honeycomb lattice and we link the optimal frequency of the synchronization to the FSA.
}\\[20pt]
\end{quote}
\end{center}
\setcounter{equation}{0}
\setcounter{figure}{0}
\setcounter{table}{0}
\setcounter{page}{1}
\setcounter{section}{0}
\makeatletter
\renewcommand{\theequation}{S\arabic{equation}}
\renewcommand{\thefigure}{S\arabic{figure}}
\renewcommand{\thepage}{S\arabic{page}}

\twocolumngrid

\section{FSA AND SCAR STABILIZATION ON THE SQUARE LATTICE}\label{Sec:FSA}

The forward scattering approximation (FSA) is crucial in the understanding of quantum scars in 1D Rydberg chains because it generates a subspace which is approximately decoupled from the rest of the Hilbert space. This implies that states spanned by the subspace thermalize very slowly \citep{Turner2017,Turner2018}. In the 1D blockade, a set of quasi-local perturbations was shown to lead to nearly perfect revivals of the maximally excited state~\citep{PhysRevLett.122.220603}. The optimal --- according to fidelity maximization --- perturbation corresponds to the perturbation which disconnect the FSA subspace almost completely from the rest of the Hilbert space. Remarkably, the optimal perturbation associates the FSA subspace to the weight vectors of an irreducible representation of an su(2) algebra. 

In this section we discuss the generalization of this mechanism to 2D lattices. First, we briefly describe the FSA algorithm for the Hamiltonian defined in Eq.~(1) of the main text and study the effects of the perturbation to the subspace. Finally we elaborate on the relation between the FSA  and su(2) algebra for the perturbations that maximize the fidelity of the state.

\subsection{FSA method}
In the following, we briefly describe the FSA method and explore the relation between the FSA and the stabilization of scars in 2D. The main idea of FSA approximation is the splitting of the Hamiltonian
\begin{equation}\label{Eq:FSAsplit}
H = H^{+}+H^{-},\quad H^{+} = (H^{-})^\dag,
\end{equation}
where $H^{+/-}$ increase/decrease the Hamming distance from the maximally excited state. The Hamming distance counts the number of spin flips required to reach the given state by the action of $H^+$ on the starting product state, $\ket{M_{A}}$. For the Hamiltonians in the form of 
\begin{equation}\label{Eq:Model_general}
H = \sum_{\vec{r}}\left({\tilde\sigma}^x_{\vec{r}}+{V}_{\textbf{r}}\right)= \sum_{\vec{r}}\tilde{\sigma}^x_{\vec{r}}\left(1+v_{\textbf{r}}\right),
\end{equation}
where  ${v}_{\vec{r}}\tilde{\sigma}^x_{\vec{r}}=V_{\vec{r}}$ as defined in Eqs.~(1)-(2) of the main text (for the square lattice), the splitting is given by
\begin{equation}\label{Eq:Hpdef}
H^{+} =  \sum_{\vec{r}\in A}\tilde{\sigma}^-_{\vec{r}}\left(1+{v}_{\textbf{r}}\right)+\sum_{\vec{r}\in B}\tilde{\sigma}^+_{\vec{r}}\left(1+{v}_{\textbf{r}}\right).
\end{equation}
We used the convention $\tilde{\sigma}^\pm_{\vec{r}} = \sigma^\pm_{\vec{r}}\prod_{\braket{\vec{r'},\vec{r}}} P_{\vec{r'}}$ for local raising and lowering operators dressed by projectors. Equation~(\ref{Eq:Hpdef}) is the same for any lattice as long as the perturbations are diagonal in the $z$-basis. The FSA subspace $\mathcal{V}$, is defined by
\begin{equation}\label{Eq:FSA}
 \text{Sp}(\mathcal{V}) = \left\{\ket{n}=\frac{(H^{+})^{n}\ket{M_{A}}}{||(H^{+})^{n}\ket{M_{A}}||}\right\}~\text{for}~n \in \{0,N\},
\end{equation} 
where $N$ is the total number of lattice sites and $\text{Sp}$ denotes the span. By definition the FSA basis vectors are orthonormal, $\braket{n_{1}|n_{2}} = \delta_{n_1,n_2}$. This subspace can be further improved by imposing an equal treatment for the two maximally excited states, $\ket{M_{A/B}}$. This can be done by generating the first $N/2$ states by using $\ket{M_{A}}$ as a vacuum and $H^+$ as a generator, and the last $N/2$ states by using  $\ket{M_{B}}$ as a vacuum and $H^-$ as a generator. We note that the states generated from the two different vacua are orthogonal to each other because they do not share the same support. The only state which requires extra care is the one whose Hamming distance from both $\ket{M_{A}}$ and $\ket{M_{B}}$ is $N/2$. For this state we pick the equal superposition $\ket{N/2+1} = H^+\ket{N/2}+H^-\ket{N/2+2}$ and normalize it to get an orthonormal basis for $\mathcal{V}$.

\begin{figure}[t]
\begin{center}
\includegraphics[width=0.7\columnwidth]{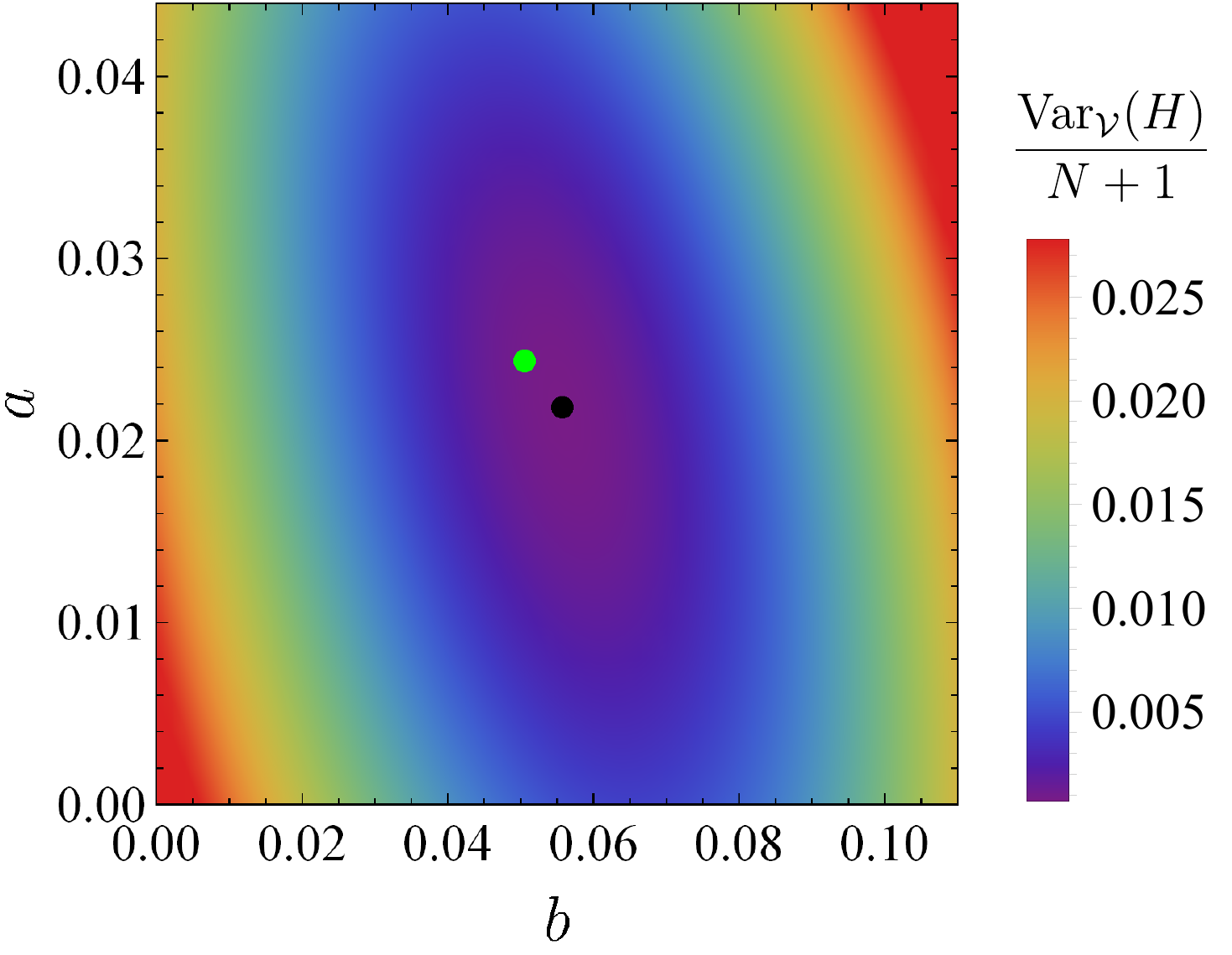}
\caption{ \label{Fig:variance} The black dot denotes the minimum of the subspace variance as a function of the perturbation strengths. The green dot shows the couplings which lead to the best revivals.}
\end{center}
\end{figure}

\begin{figure}[t]
\begin{center}
\includegraphics[width=0.99\columnwidth]{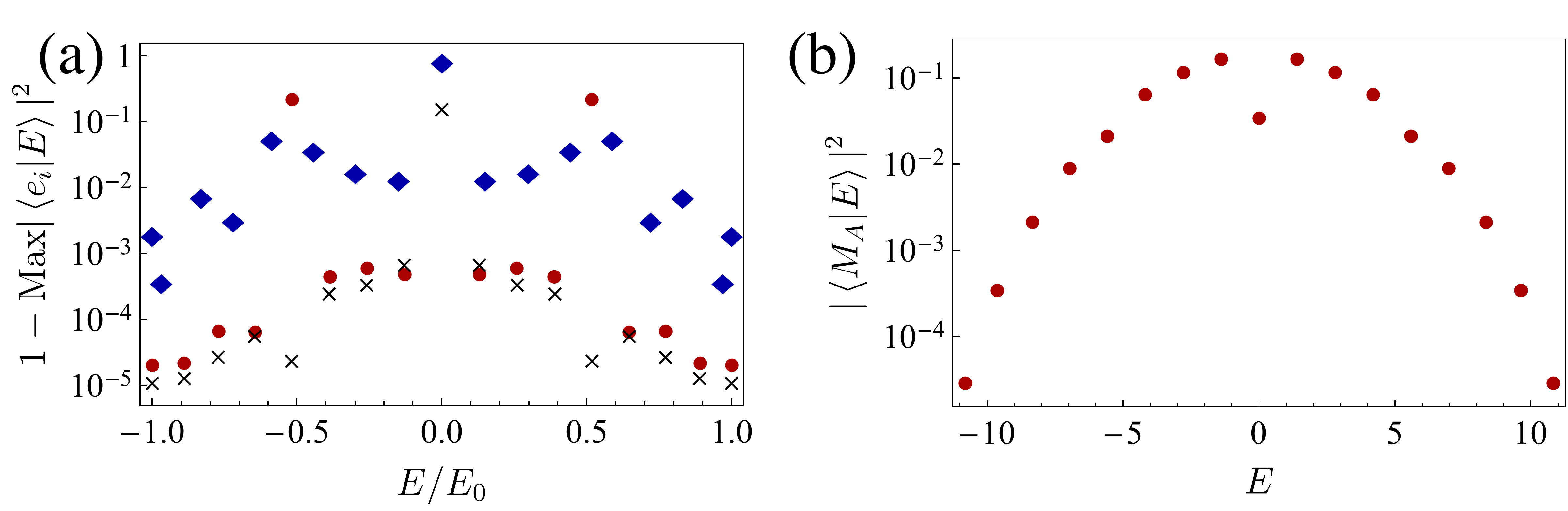}
\caption{ \label{Fig:Overlaps} (a) The comparison of the overlaps between the eigenstates of the projected Hamiltonian $Q_{\mathcal{V}}H Q_{\mathcal{V}}$ and the full Hamiltonian $H$ demonstrates the  improved accuracy of the approximate eigenstates for the optimal perturbation. The blue diamonds denote the unperturbed model, the red circles denote the perturbation which maximizes fidelity, and the black crosses denote the perturbation that minimizes subspace variance, Eq~(\ref{Eq:Var}). (b) Overlap of the maximally excited state with the eigenstates of the full Hamiltonian for the perturbation that maximizes fidelity. Only the overlaps with the eigenstates that are approximated by the FSA subspace are shown.}
\end{center}
\end{figure}
To measure how disconnected the subspace is, we use the subspace variance,
\begin{equation}\label{Eq:Var}
\text{Var}_{\mathcal{V}}(H) = \text{Tr}(Q_{\mathcal{V}} H^2)-\text{Tr}(Q_{\mathcal{V}} H)^2,
\end{equation}
where $Q_{\mathcal{V}} = \sum_{n}\ket{n}\bra{n}$ is the projector to the subspace $\mathcal{V}$. Eq.~(\ref{Eq:Var}) can be qualitatively understood by realizing that its minimum value, zero, is achieved only if $\dim(\mathcal{V}) = N+1$ eigenstates of the Hamiltonian are fully supported in $\mathcal{V}$. In this case the dynamics of any state that is a superposition of those eigenstates can not lead to thermalization. The subspace is constructed in such a way to trivially include $\ket{M_{A}} = \ket{0}$. For this reason, if $\text{Var}_{\mathcal{V}}(H)$ is very small, the state $\ket{M_{A}}$ is expected to thermalize very slowly. 

For the rest of the section we focus on the $4 \times 4$ square lattice and the perturbation defined in Eqs.~(2)-(3) of the main text. Similar results hold for the honeycomb lattice described in the next section and are expected to hold for various different bipartite lattices. In Figure~\ref{Fig:variance} we show the dependence of the subspace variance, $\text{Var}_{\mathcal{V}}(H)$, on the perturbation parameters. We observe that the minimum $V_{H}=(a_{H},b_{H}) \approx (0.0217,0.0556)$ is close to the value that reproduces the best fidelity revivals $V_{c}=(a_{c},b_{c}) \approx (0.0244,0.0506)$. 

To further quantify the effect of the perturbations, we compare the eigenstates of the full Hamiltonian, $\{\ket{E}\}$, to those of the Hamiltonian projected to the
 subspace, $\{\ket{e}\}$. In Figure~\ref{Fig:Overlaps}(a) we show the highest overlap between each approximate eigenstate and the eigenstates of the full model, $\text{Max}_{\ket{E}}|\braket{e_{i}|E}|^2$. As expected, the approximate eigenstates of the deformed model are closer to the exact eigenstates. On the other hand there is little difference between the  fidelity optimized perturbation, $V_{c}$, and the variance optimized perturbation, $V_{H}$. 
 
 The disagreement between $V_{H}$ and $V_{c}$ can be understood from Fig.~\ref{Fig:Overlaps}(b). The overlap of the maximally excited state is much larger for the two eigenstates close to the middle of the spectrum, which means that the dynamics of $\ket{M_{A}}$ is mostly sensitive to those two eigenstates and not to the rest of the subspace $\mathcal{V}$. Indeed from Figure~\ref{Fig:Overlaps}(a), we observe that the two middle eigenstates are closer to the approximate eigenstates for the perturbation that maximizes fidelity, $V_{c}$. To verify that observation we find that the perturbation which minimizes the energy variance of only the two relevant eigenstates, $(a_{0},b_{0}) \approx (0.022,0.054)$ is much closer to $V_{c}$.
\begin{figure}[t]
\begin{center}
\includegraphics[width=0.99\columnwidth]{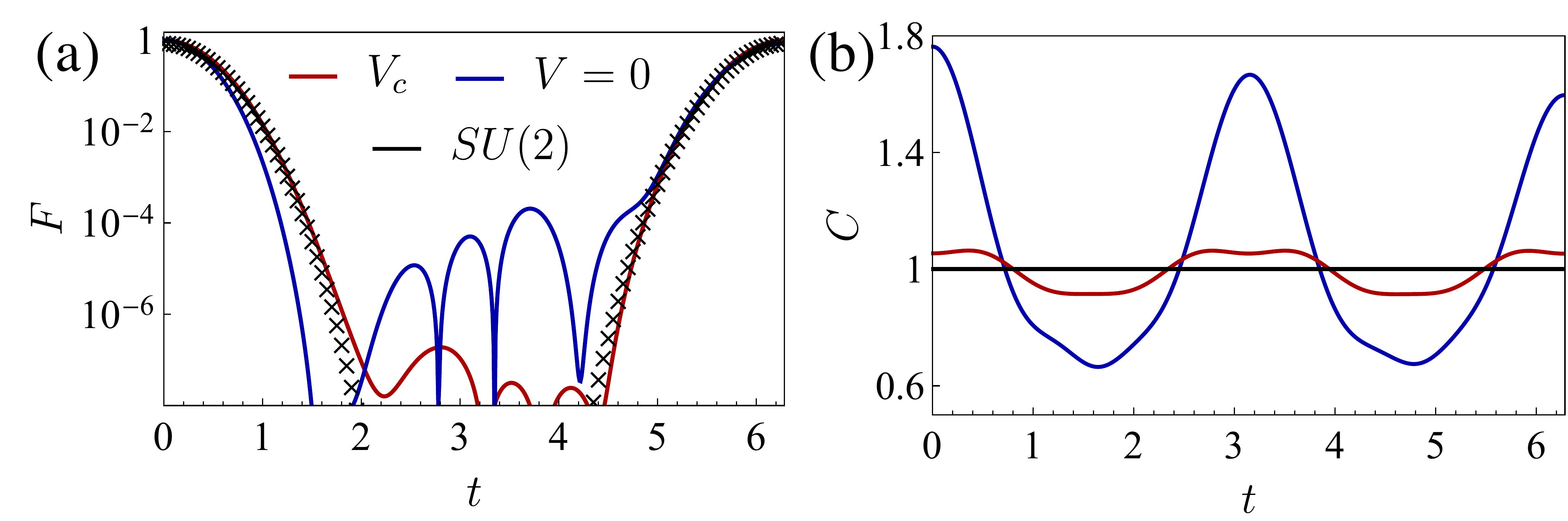}
\caption{ \label{Fig:su2_dynamics} Dynamics for a quench from the highest weight vector $\ket{M_{A}}$. (a) Fidelity dynamics for the unperturbed and optimally  perturbed ($V_{c}$) Hamiltonians in comparison to the fidelity revivals of the highest weight vector of an $N/2$-dimensional $\mathrm{su(2)}$ algebra. (b) Dynamics of the normalized Casimir operator.
 }
\end{center}
\end{figure}

\subsection{FSA and su(2) algebra}

In this section we show that the FSA generators,  $\{H^+,H^-,H^z= [H^+,H^-]\}$, for the optimal perturbation, also approximately generate an $N/2$-dimensional $\mathrm{su(2)}$ representation.  The orthonormal basis of the subspace $\mathcal{V}$,  $\{\ket{n}\}$, form the weight vectors of the representation with the highest/lowest weight vectors being $\ket{M_{A/B}}$. We note that the generators do not act in a special way outside of $\mathcal{V}$.

According to these definitions, the Hamiltonian -- if applied in the subspace -- just rotates the $N/2$-dimensional spin around the $x$-axis, $H = H^{x} = H^{+}+H^{-}$. To verify these assumptions we compare the exact dynamics to the dynamics of an exact $N/2$-dimensional su(2) algebra $\{S^+,S^-,S^z= [S^+,S^-]\}$. We choose the representation where the period  of the quench from the highest weight state, $e^{i S^{x} T_{SU(2)}}\ket{M_{A}}=\ket{M_{A}}$, is just  $T_{SU(2)}= 2 \pi$. The FSA algebra is renormalized in order for the period of the fidelity revivals to be $T = T_{SU(2)}$:
\begin{equation}
 H^{\pm} \rightarrow \frac{H^{\pm}}{2 \pi T}, \; H^{z} \rightarrow \frac{H^{z}}{(2 \pi T)^2}, 
\end{equation}
where $T$ is the non-normalized period of revivals in the quantum system. The (normalized) Casimir operator of the algebra is
\begin{equation}
C = \frac{2\{H^+,H^-\} +4(H^z)^2}{\frac{N}{2}\left(\frac{N}{2}+1\right)}.
\end{equation}
If the FSA subspace generators form an irreducible representation of an $N/2$-dimensional $\mathrm{su(2)}$ algebra, the Casimir operator will be an integral of motion, $C(t) = 1$. 
In Figure~\ref{Fig:su2_dynamics} we compare the dynamics in a quench from the maximally excited state with the $\mathrm{su(2)}$ algebra prediction. The fidelity of the perturbed $(V_{c})$ Hamiltonian agrees well with the $\mathrm{su(2)}$ prediction. On the contrary, the unperturbed system deviates significantly as the dynamics move the state further from $\ket{M_{A}}$. In the same spirit, the Casimir operator in the stabilized model oscillates with a much smaller amplitude compared to the unperturbed Hamiltonian. 

In addition to the dynamics, the matrix elements  $\braket{n|H^{z}|n}$ (not shown), indicate that the optimal perturbation indeed makes the FSA subspace more $\mathrm{su(2)}$-like. These results are reminiscent of the (almost) perfect scars of the one dimensional Rydberg blockades~\citep{PhysRevLett.122.220603}. We believe that the stabilizing perturbation in this work is the higher dimensional analog of the range-1 perturbation found in~\citep{Khemani2018} and extended in~\citep{PhysRevLett.122.220603}. The difference in higher dimensional models is that the number of possible perturbation terms rapidly increases with the range of the deformation. This fact along with the limited range of system sizes in 2D makes the search for longer range perturbations very challenging.

\section{STABILIZATION OF THE HONEYCOMB LATTICE}

In this section we show that the stabilizing perturbations used in the main text can be generalized to different bipartite lattices. As an example we study the Rydberg blockade in the honeycomb lattice. The Hamiltonian of the system is
\begin{equation}\label{Eq:Model}
H = \sum_{\vec{r}}\sigma^x_{\vec{r}}\left(1+{v}_{\textbf{r}}\right)\prod_{\braket{\vec{r'},\vec{r}}} P_{\vec{r'}},\quad {v}_{\vec{r}}  = (a \mathcal{P}_{\vec{r}}^{l}+ b\mathcal{P}_{\vec{r}}^{2}).
\end{equation}
where the perturbations are defined as
\begin{subequations}\label{Eq:perturbationsSUP}
\begin{eqnarray}
\mathcal{P}^{l}_{\vec{r}} &=& \sum_{\braket{\braket{\vec{r'},\vec{r}}}}P_{\vec{r'}},\label{Eq:p1}\\
\mathcal{P}^{2}_{\vec{r}} &=& \sum_{\braket{\braket{(\vec{r'},\vec{r''}),\vec{r}}}}P_{\vec{r'}}P_{\vec{r''}},\label{Eq:p2}
\end{eqnarray}
\end{subequations}
The projector $\mathcal{P}^{l}_{\vec{r}}$ is a superposition of projectors acting on the next nearest neighbors sites to the site at $\vec{r}$. The projector $\mathcal{P}^{2}_{\vec{r}}$ is a superposition of the projectors, simultaneously applied to all next nearest neighbor sites to $\vec{r}$ which share a common neighbor with the site at $\vec{r}$. These projectors are analogs of the projectors $\mathcal{P}^{l},\mathcal{P}^{3}$ of the square lattice given in the main text. The main difference in the honeycomb, is the absence of next nearest neighbors which share two common neighbors with the site at $\vec{r}$ and thus the analog of the projector $\mathcal{P}^{d}$ is not needed.

The honeycomb lattice has a two atoms in the unit cell that correspond to partitions $A/B$. The states which have the maximum number of excitations  $\ket{M_{A/B}}$ corresponds to the states where the sites in sublattice $A/B$ are excited. As in the square lattice we  maximize the fidelity ${F(t)=|\langle M_A|e^{-iHt}|M_A \rangle|^2}$, at the instance of the first revival at time $T$. The results are summarized in the following table: 
\begin{center}
\begin{tabular}{c c c c   }
\; System size  \;  &a &b &\; $-1/N\ln F(T)$ \;\\
  \hline
 $2\cdot 3 \times 3$   & 0.03038  \;&\; 0.06345 \;& $3.83 \times 10^{-5}$ \\
 $2\cdot 4 \times 4$   & 0.03037  \;&\; 0.06203 \;& $1.87 \times 10^{-5}$ \\
\end{tabular}
\end{center}
where we considered lattices that consist of $N=18$ sites ($3\times 3$ unit cells) and $N=32$ sites ($4\times 4$ unit cells).
The optimal perturbation is almost independent of the system size, which means that the stabilization is not a result of the small size of the Hilbert space.
\begin{figure}[t]
\begin{center}
\includegraphics[width=0.99\columnwidth]{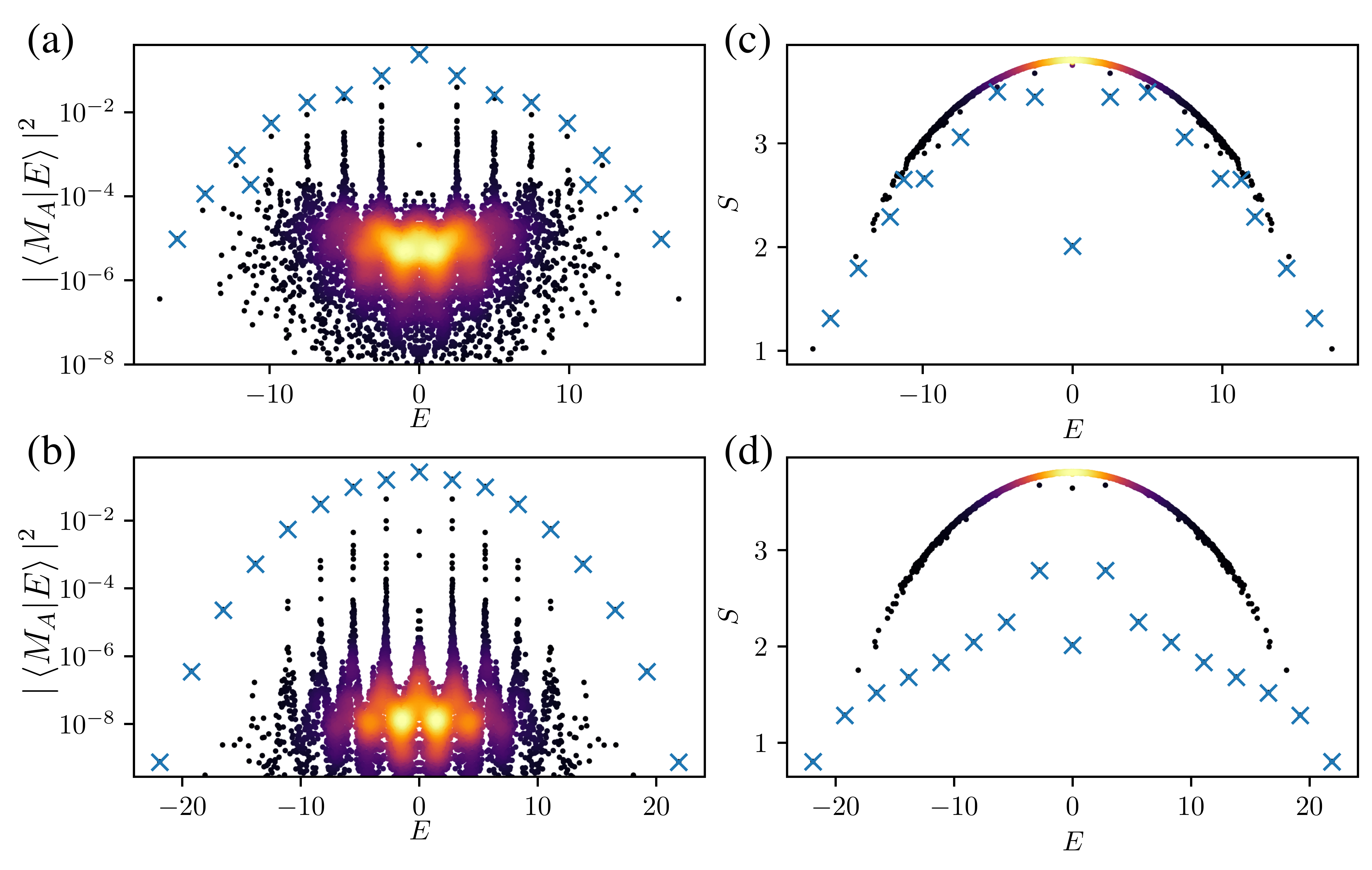}
\caption{ \label{Fig:honeycomb} (a,b) Overlaps between the maximally excited state and the eigenstates of the unperturbed (a), and the optimally perturbed hamiltonian (b). (c) Entanglement entropy for a subsystem of size $4 \times 2$ for the unperturbed, and (d) the optimally perturbed hamiltonian . The crosses label the eigenstates which have highest overlaps with the maximally excited states at a given energy density. }
\end{center}
\end{figure}
In Figure~\ref{Fig:honeycomb} we compare the unperturbed and the stabilized Hamiltonians for a  lattice with $N=32 $ sites with PBC. We observe that the entanglement entropy of a bipartition of the honeycomb lattice has similar features to the square lattice, i.e.~the stabilization of the scars forms a band of slightly entangled eigenstates. The same phenomenon appears in the overlaps between the eigenstates and the maximally excited state  $|\braket{E|M_{A}}|^2$. There is a single eigenstate with much higher overlap than all other eigenstates in the same energy density. 

\begin{figure}[t]
\begin{center}
\includegraphics[width=0.99\columnwidth]{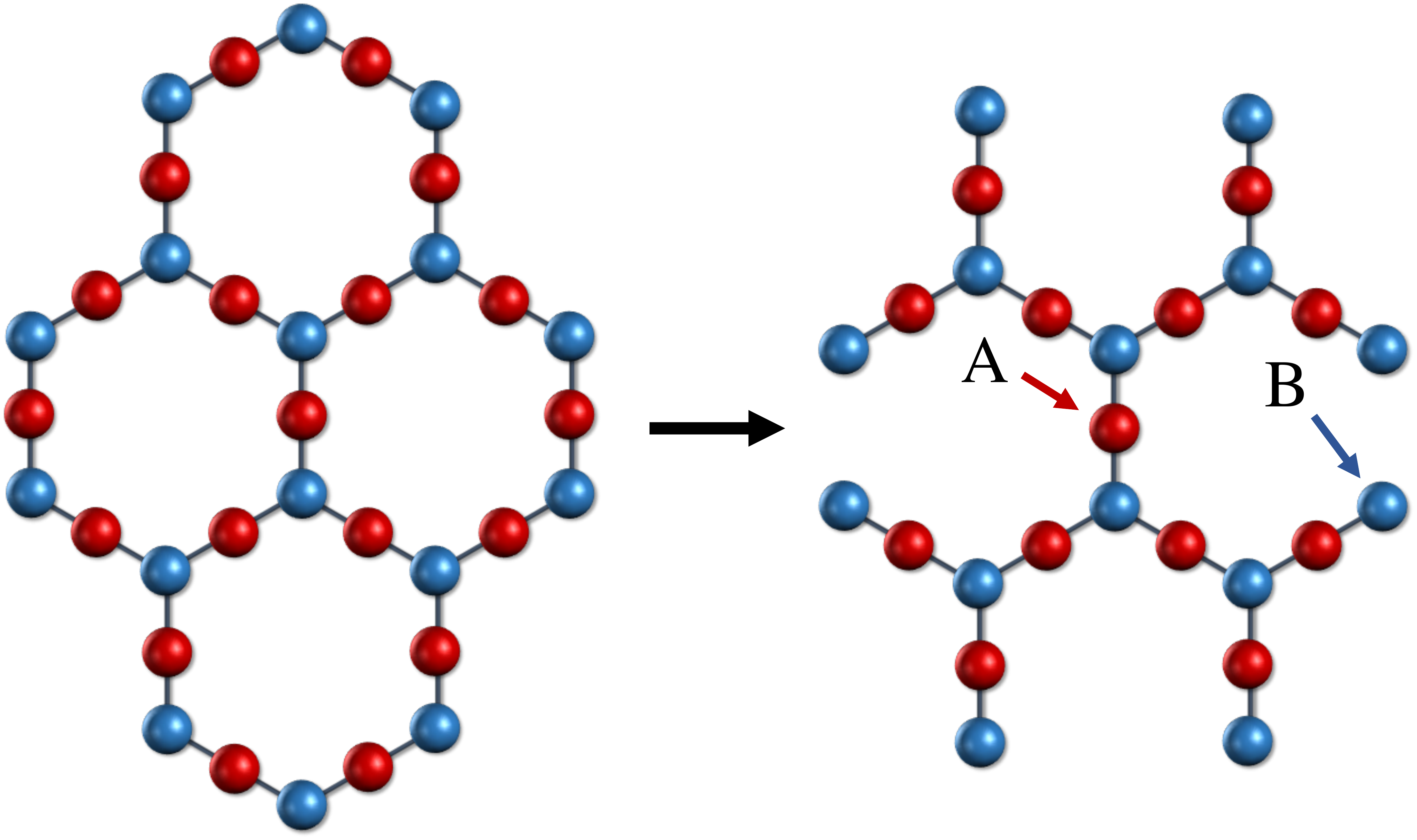}
\caption{ \label{Fig:tree} Mapping of the decorated lattice onto a decorated Cayley tree. The wavefunction of the decorated tree tensor network combines local tensors with two virtual indices which are attached onto the links of the network (sublattice $A$) and local tensors with three virtual indices which are attached to the vertices of the network (sublattice $B$).}
\end{center}
\end{figure}
\section{DECORATED LATTICE}
In this section we first describe the TTS ansatz used in the main text and compare the approximate dynamics to the exact dynamics of the decorated lattice. We then describe how to generalize the FSA method to such decorated lattices and show how to estimate the optimal Rabi frequency, $\omega_{c}$, by requiring the FSA subspace $\mathcal{V}$ to be symmetric. 
\subsection{TTS approximation to decorated lattices}
To study the decorated lattice, we use a decorated Cayley tree lattice,  Figure~\ref{Fig:tree}. Tree tensor networks can accurately approximate the properties of the ground states~\citep{li2012efficient} and in certain cases the dynamics~\citep{michailidis2019slow} of lattices with loops. For the decorated tree we use the mean field-like variational wave function,
\begin{equation}\label{Eq:state}
\begin{split}
&\ket{\psi} =  M\sum_{N_{A}N_{B}} c^{N/2-N_{A}}_{\downarrow_{A}} c^{N_{A}}_{\uparrow_{A}} c^{L/2-N_{B}}_{\downarrow_{B}} c^{N_{B}}_{\uparrow_{B}}\ket{N_{A},N_{B}}, 
\end{split}
\end{equation}
where $\ket{N_{A},N_{B}}$ is an equal superposition of the many body states in the computational basis where the number of excitations in each subblatice is fixed, ${\sum_{j\in A/B}n_{j} = N_{A/B}}$. To have a normalized state in the thermodynamic limit we use the following parametrization of the weights, 
\begin{subequations}\label{Eq:weights}
\begin{eqnarray}
c_{\downarrow_{A}} &=& \cos{\theta_{A}},\quad c_{\uparrow_{A}} =ie^{-i \phi_{A}}\tan{\theta_{B}},\\
c_{\downarrow_{B}}&=& \cos{\theta_{B}},\quad c_{\uparrow_{B}} =ie^{-i \phi_{B}}\tan{\theta_{A}}.
\end{eqnarray}
\end{subequations}
The ansatz of states defined by Eqs. (\ref{Eq:state})-(\ref{Eq:weights}) is identical to the tensor tree ansatz (TTS) 
\begin{equation}\label{Eq:TTS_full}
\ket{\psi} = \text{Tr}\prod_{i\in A}\mathcal{A}^{n_{i}}\ket{n_{i}}\prod_{j\in B}\mathcal{B}^{n_{j}}\ket{n_{j}},
\end{equation}
where the trace is taken over all physical indices $n_{i},n_{j}$ and over the virtual dimensions of the local tensors in the appropriate order to reproduce the TTS. The tensors $\mathcal{A}, \mathcal{B}$ are defined as,
\begin{equation}\label{Eq:Tensors}
\begin{split}
 &O^{\downarrow}_{a_{1},\dots a_{c_{i}-1},0}(i) = \cos^{I}{\theta_{i}}\sin^{c_{i}-1-I}{\theta_{i}},\\
&O^{\uparrow}_{0,\ldots 0,1}(i) = i e^{-i \phi_{i}},\\
&\mathcal{A}\equiv O(i = A)  ,\; \mathcal{B}\equiv O(i = B)   \\
\end{split}
\end{equation}
where $c_{i}$ is the connectivity the sublattice $A/B$ and $I  = \sum_{l=1}^{c_{i}-1}a_{l}$ counts the number of ``excitations" on the $c_{i}-1$ adjacent sites. The difference from Ref.~\citep{michailidis2019slow} is that the network contains two types of tensors with different number of virtual indices corresponding to sublattices $A/B$. The  tensor $\mathcal{A}$ has two virtual indices (similar to local tensors in a matrix product state) while the local tensor $\mathcal{B}$ has three virtual indices. 

The state where sites in sublattice $A$ are excited and sites in sublattice $B$ are in the ground state is labeled $\ket{M_{A}}$ and the state where the sites in sublattice $B$ are excited and the sites in sublattice $A$ are in the ground state is labeled $\ket{M_{B}}$. From Eq.~(\ref{Eq:weights}), we observe that  $\ket{M_{A}}$ corresponds to the solutions of $\{\cos{\theta_{A}}= 0,\cos{\theta_{B}}= 1\}$ and $\ket{M_{B}}$ to the solutions of $\{\cos{\theta_{B}}= 0,\cos{\theta_{A}}= 1\}$. In both cases we can fix $\phi_{A}=\phi_{B}=0$.  We note that for the decorated lattice the wave function is not normalized for the states $\ket{M_{A}},\ket{M_{B}}$. We can deal with this problem by either initializing the system in a state which is $\epsilon \ll 1$ away from the singular states or by regularizing the equations of motion.
\begin{figure}[t!]
\begin{center}
\includegraphics[width=0.99\columnwidth]{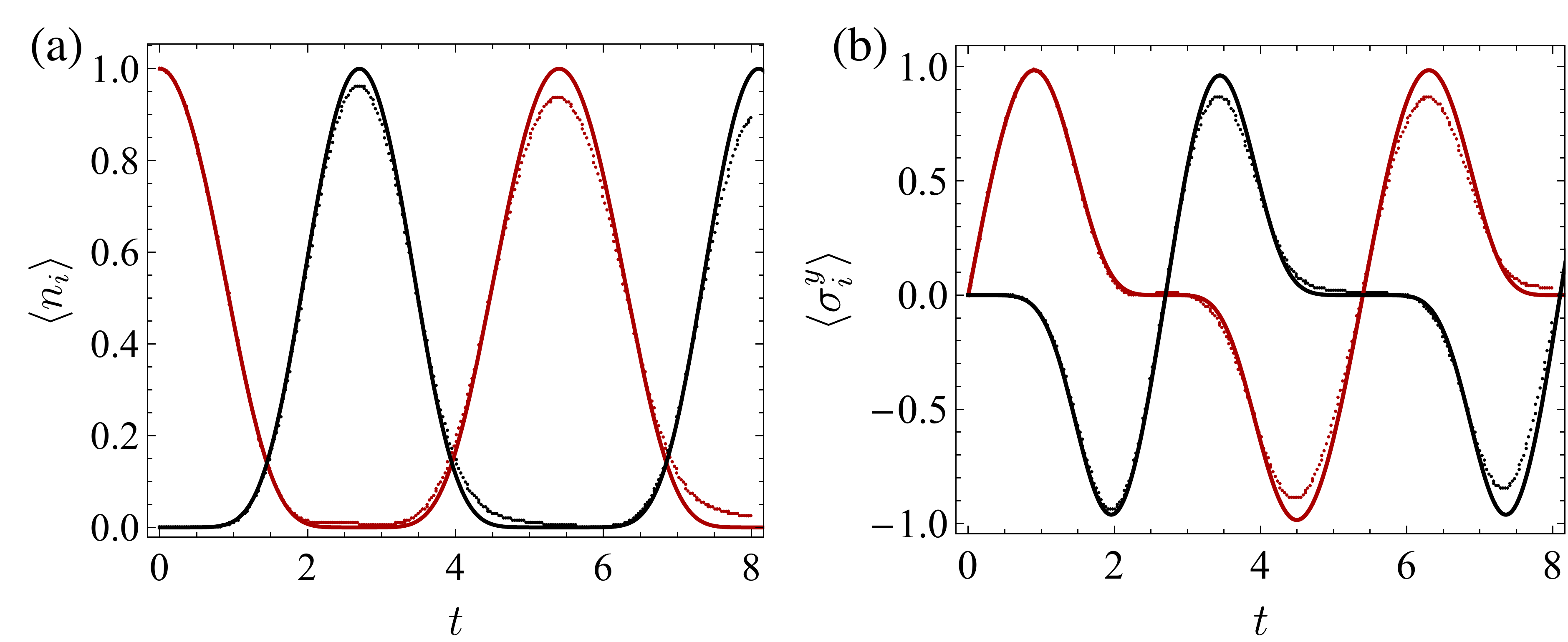}
\caption{ \label{Fig:localops} Comparison of the dynamics of local observables using exact evolution in the decorated lattice (dots) and TDVP on the decorated tree lattice (lines). The red/black colors denotes the dynamics of local observables acting on $A/B$ subblattice. The initial state is $\ket{M_{A}}$ and $\omega = \omega_{c}$. (a) Evolution of the local excitation density and (b) $\sigma^y$ expectation value. }
\end{center}
\end{figure}
The dynamics of the system is generated by the Hamiltonian defined in Eq.~(4) of the main text. The time-dependent variational principle (TDVP)~\citep{kramer1981geometry,Haegeman,michailidis2019slow}, is employed to find the equations of motion in Eq.~(5) of the main text. To avoid singularities we regularize the tangent function as described in the main text. In Figure~\ref{Fig:localops} we compare the dynamics of local observables using the exact evolution for a system with $N = 20 $ sites ($2 \times 2$ unit cells with PBC)  and the dynamics generated by the TDVP, Eq.~(\ref{Eq:EOM}). The local correlation functions agree well within a period of the classical orbit. This implies that all physical processes associated to the coherent dynamics are well captured by our ansatz.  
\subsection{FSA for the decorated lattice}
\begin{figure}[t!]
\begin{center}
\includegraphics[width=0.99\columnwidth]{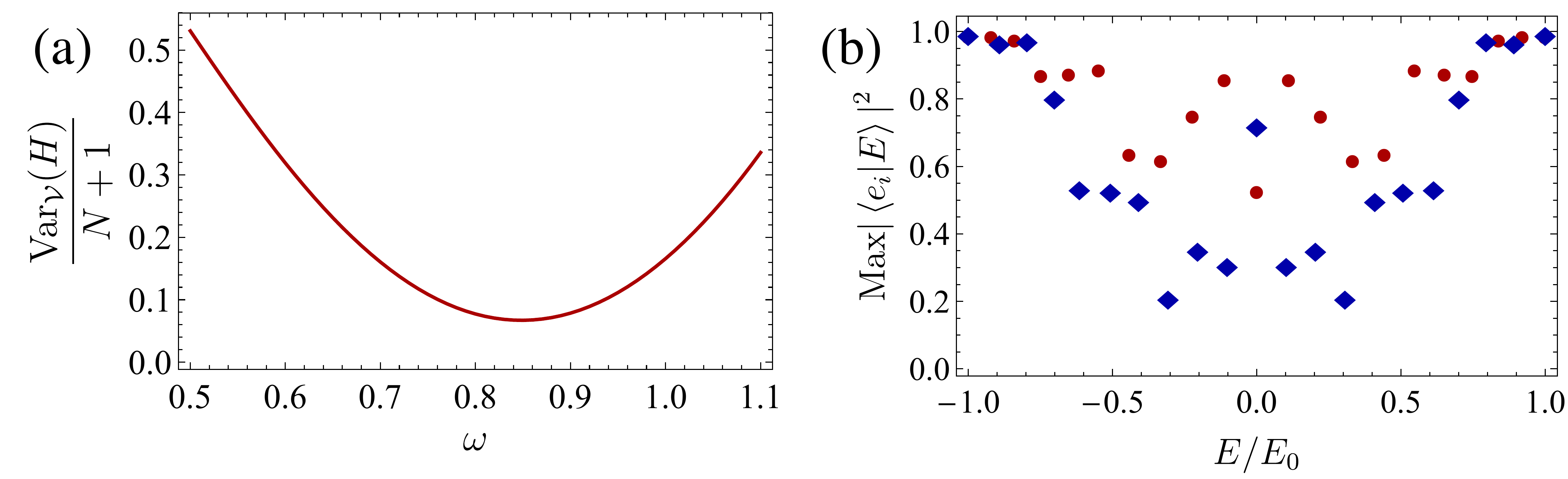}
\caption{ \label{Fig:dec_FSA} (a) Subspace variance of the FSA subspace as a function of frequency detuning. (b) Maximum overlap between each FSA eigenmodes and the eigenstates of the Hamiltonian. The red color denotes $\omega = \omega_{0}$ and the blue color by $\omega = 1$.  }
\end{center}
\end{figure}
In this section we extend the FSA method described in Sec.~\ref{Sec:FSA} to decorated lattices and show that the optimal frequency can be obtained by imposing a condition of symmetry between $\ket{M_{A,B}}$ states within the FSA subspace. The splitting of the Hamiltonian is performed according to Eq.~(\ref{Eq:FSAsplit}). The two ladder operators read:
\begin{equation}\label{Eq:decor}
H^{\pm}= \omega \sum_{\vec{r}\in A}\tilde{\sigma}^{\mp}_{\vec{r}} + \sum_{\vec{r} \in B}\tilde{\sigma}^{\pm}_{\vec{r}}.
\end{equation}
The complication in the decorated lattice is hidden in the unequal number of excitations in $\ket{M_{A}}$ and $\ket{M_{B}}$. For $N$ sites, $\ket{M_{A}}$ contains $N^\text{max}_{A}=(3/5)N$ excitations while $\ket{M_{B}}$ contains $N^\text{max}_{B}=(2/5)N$ excitations. We chose to generate the subspace by treating $\ket{M_{A/B}}$ on equal ground using the same basic idea as for the square lattice,
\begin{equation}
\begin{split}
&\text{Sp}(\mathcal{V}_{A}) = \left\{\ket{n}=(H^{+})^{n}\ket{M_{A}}\right\},~n \in \{0,\frac{3N}{5}-1\} \\
 & \text{Sp}(\mathcal{V}_{B}) = \left\{\ket{N+1-n}=(H^{-})^{n}\ket{M_{B}}\right\},~n \in \{0,\frac{2N}{5}-1\}\\
  &\ket{3N/5+1} = H^{+}\ket{3N/5}+H^{-}\ket{3N/5+2},\\
  &\text{Sp}(\mathcal{V}) = \text{Sp}(\mathcal{V}_{A}) \cup \text{Sp}(\mathcal{V}_{B})\cup  \ket{3N/5+1}.
\end{split}
\end{equation}
We note that the basis of $\mathcal{V}$ defined above, ${\{\ket{n}\},\; n \in \{0,N\}} $, is orthogonal. For the rest we use a normalized basis $\ket{n}\rightarrow\ket{n}/\norm{\ket{n}}$. 

The first difference of the FSA applied to the decorated lattice instead of the normal lattice is that we generate more states by applying $H^{+}$ to $\ket{M_{A}}$ than by applying $H^{-}$ to $\ket{M_{B}}$. However this asymmetry is not a priori problematic. The second difference, which proves to be much more sensitive to the fidelity of the revivals can be quantified by the difference in the norms,  
\begin{equation}\label{Eq:assymetryfull}
\norm{(H^{+})^{n}\ket{M_{A}}}\neq \norm{(H^{-})^{n}\ket{M_{B}}}, \; \forall n \in \{1,2 N/5\}.
\end{equation}
A particular choice of $\omega$ that almost fixes this asymmetry for all $n$ is found by equating the norms for $n=1$,
\begin{equation}\label{Eq:assymetry}
\norm{H^{+}\ket{M_{A}}} = \norm{H^{-}\ket{M_{B}}} \Rightarrow \omega \sqrt{\frac{3N}{5}} = \sqrt{\frac{2N}{5}}.
\end{equation}
The solution $\omega_{s} = \sqrt{{2}/{3}}\approx 0.81$ is practically the same as the frequency that produces the optimal fidelity, ${\omega_{c} \approx 0.8}$. Even though we do not have an analytic understanding behind this agreement, this criterion can be understood via properties of the quantum trajectory that passes through $\ket{M_{A/B}}$ states. By equating the norms in Eq.~(\ref{Eq:assymetry}) the physical property that emerges in the quantum trajectory is that the rates ${\norm{\frac{d\ket{M_{B}}}{dt}}=\norm{\frac{d\ket{M_{A}}}{dt}}}$. In the same spirit since the  norms for $n\neq 1$ are approximately equal by that choice of $\omega$, higher time derivatives of the two states are approximately equal in magnitude. This result can be generalized as $\omega = \sqrt{{c_{A}}/{c_{B}}}$ for decorated lattices with different sublattice connectivities $c_{A},c_{B}$. 

In Figure~\ref{Fig:dec_FSA}(a) we numerically justify the value of $\omega$ by evaluating the subspace variance, Eq.~(\ref{Eq:Var}), for different detunings. The optimal value $\omega_{0} \approx 0.84$ is very close to the exact result and the same as the TDVP result up to the first two significant digits. We don't have an understanding for the small difference between $\omega_{s}$ and $\omega_{0}$. It could possibly be due to finite size effects. The improvement of quality of the revivals is further justified by the maximum overlaps of the eigenstates of the Hamiltonian projected in the FSA subspace and the exact eigenstates of the system, Figure~\ref{Fig:dec_FSA}(b).

\end{document}